\documentstyle[jlp,twoside]{article}

\newcommand{\hd}{head}
\newcommand{\pb}{positivebody}

\newcommand{\lar}{\leftarrow}

\newcommand{\n}{\mbox{\bf{not}}}

\newcommand{\at}{\hbox{\it At}}

\newcommand{\smn}{\setminus}
\newcommand{\la}{\langle}
\newcommand{\ra}{\rangle}
\newcommand{\St}{\mbox{{\it ST}}}

\begin{document}

\main{EXTREMAL PROBLEMS IN LOGIC PROGRAMMING AND STABLE MODEL
COMPUTATION}{PAWEL CHOLEWINSKI
AND 
MIROSLAW TRUSZCZYNSKI}
\affil{Miroslaw Truszczynski, Computer Science Department,
University of Kentucky, Lexington, KY 40506-0046, USA,
{\tt mirek@cs.engr.uky.edu}; Pawel Cholewinski, HyBrithms Corporation,
10632 NE 37th Circle, Blgd. \#23, Kirkland, WA 98033, {\tt
pch@hybrithms.com}\\ \ }

\begin{abstract}
We study the following problem: given a class of logic programs 
$\cal C$, determine the maximum number of stable models of a program 
from $\cal C$. We establish the maximum for the class of all logic 
programs with at most $n$ clauses, and for the class of all logic 
programs of size at most $n$. We also characterize the programs for
which the maxima are attained. We obtain similar results for the class
of all disjunctive logic programs with at most $n$ clauses, each of 
length at most $m$, and for the class of all disjunctive logic programs 
of size at most $n$. Our results on logic programs have direct 
implication for the design of algorithms to compute stable models. 
Several such algorithms, similar in spirit to the Davis-Putnam procedure, 
are described in the paper. Our results imply that there is an algorithm
that finds all stable models of a program with $n$ clauses after
considering the search space of size $O(3^{n/3})$ in the worst
case. Our results also provide some insights into the question of
representability of families of sets as families of stable models
of logic programs.
\end{abstract}

\section{INTRODUCTION}
\label{intro}

In this paper we study extremal problems appearing in the context
of finite propositional logic programs. Specifically, we consider 
the following problem: given a class of logic programs $\cal C$, 
determine the maximum number of stable models a program in 
$\cal C$ may have. 
Extremal problems have been studied in other disciplines, 
especially in combinatorics and graph theory \cite{bo78}. However, 
no such results for logic programming have been known so far.

We will consider finite propositional disjunctive logic programs 
built of {\em clauses} ({\em rules}) of the form
\[
a_1\vee \ldots \vee a_k \lar b_1,\ldots , b_m,\n(c_1),\ldots,\n(c_n),
\]
where $a_i$, $b_i$ and $c_i$ are atoms. In an effort to establish a
semantics for disjunctive logic programming, Gelfond and Lifschitz 
\cite{gl90a} introduced the notion of an {\em answer set} of 
a disjunctive program. It is well-known that for {\em normal} logic
programs (each clause has exactly one literal in the head), answer sets
coincide with {\em stable} models \cite{gl88,gl90a}. We will denote 
the set of answer sets of a disjunctive program $P$ (stable models, 
if $P$ is normal) by $\St(P)$ and we will set
\[
s(P) = |\St(P)|.
\]

Given a class $\cal C$ of disjunctive programs, our goal will be 
to determine the value of
\[
\max\{s(P)\colon P\in {\cal C}\}.
\]
We will also study the structure of {\em extremal} programs in 
$\cal C$, that is, those programs in $\cal C$ for which the maximum
is attained.

We will focus our considerations on the following classes of programs:
\begin{enumerate}
\item ${\cal DP}_{n,m}$ --- the class of disjunctive programs with at
most $n$ clauses and with the length of each clause bounded by $m$
\item ${\cal LP}_n$ --- the class of normal logic programs with
at most $n$ clauses.
\end{enumerate}
We will establish the values
\[
s(n) = \max\{s(P)\colon P\in {\cal LP}_n\}
\] 
and 
\[
d(n,m) = \max\{s(P)\colon P\in {\cal DP}_{n,m}\}.
\]
We will show that $s(n) = \Theta(3^{n/3})$ (an exact formula will be
given) and $d(n,m) = m^n$,
and we will characterize the corresponding {\em extremal} programs. 

We will also show that the bound for logic programs can be improved if
additional restriction on the length of a clause is imposed. We will
study the class ${\cal L}_n^2$ of logic programs with $n$ clauses such
that each clause has at most one literal in its body. We will show that
if $P$ is in ${\cal L}_n^2$, then $s(P) = O(2^{n/4})$.

We will also study classes of programs defined by imposing
restrictions on the total size of programs. By the {\em size} of a
program $P$, we mean the total number of atom occurrences in $P$.
We will investigate the following classes of programs:
\begin{enumerate}
\item ${\cal DP}_{n}$ --- the class of disjunctive programs with 
size at most $n$
\item ${\cal LP}'_n$ --- the class of normal logic programs with
size at most $n$
\end{enumerate}
and obtain similar results to those listed above.

The motivation for this work comes from several sources. 
First of all, this work has been motivated by our
efforts to develop fast algorithms for computing stable models of logic
programs. It turns out that bounding the number of stable models and
search for extremal logic programs are intimately connected to some
recursive algorithms for computing stable models. Two results
given in Section \ref{norm} (Corollaries \ref{cor.a} and \ref{cor.r})
imply both the bounds on the number of
stable models, and a whole spectrum of algorithms to compute 
stable models. These algorithms share some common features with
the Davis-Putnam procedure for testing satisfiability of CNF formulas.
One of these algorithms is similar to the algorithms
recently described and studied in \cite{vsnv95,nie95,ns95}. The corollaries 
also imply the worst-case bounds on the size of the
search space traversed by those algorithms. 

Let us note here that in order to lead to implemented systems for 
computing stable models, several research issues remain to be resolved. 
In particular, heuristics for choosing atoms and rules in the algorithms
presented in Section \ref{algs} must be studied. Simlarly, the effects of
using well founded semantics as a preprocessing mechanism, which is
known to be critical for the performance of the s-models system
\cite{ns96}, has to be investigated. Finally, in order to gain actual
insights into the quality of the algorithms proposed here and compare
them to other systems (such as s-models), extensive experimental studies 
is necessary. All these issues are the subject of our current studies.

Additional motivation for our work presented here comes from 
considerations of expressive power of logic programming and 
of representability issues. Both concepts help
understand the scope of applicability of logic programming as a 
knowledge representation tool. Disjunctive logic programs with answer set
semantics (logic programs with stable model semantics) can be viewed 
as encodings of families of sets, namely, of the families of their 
answer sets (stable models). 
A family of sets ${\cal F}$ is {\em representable} if 
there is a (disjunctive) logic program $P$ such that
\[
\St(P) = {\cal F}.
\]
Important problems are: (1) to find properties of representable
families of sets, and (2) given a representable family of sets $\cal F$, to
find possibly concise logic program representations of $\cal F$. Related
problems in default logic have been studied in \cite{mtt96}.
It is well-known \cite{gl90a} that every representable family of sets
must be an antichain. Our study of extremal problems in logic 
programming provide additional conditions. Namely, every family of sets
representable by a program from ${\cal DP}_{n,m}$ must have cardinality
bounded by $m^n$ and every family of sets representable by a logic program from
${\cal LP}_n$ must have size bounded by $3^{n/3}$. 
The best bound known previously for families 
of sets representable by logic programs from ${\cal LP}_n$ was
$\approx 0.8\times 2^n/\sqrt{n}$.

In addition, the results of this paper allow some comparison of
the expressive power of different classes of programs. For example,
there is a disjunctive logic program of size $n$ with $\Theta(2^{n/2})$
answer sets while the largest cardinality of a family of sets 
representable by a logic program of size $n$ is only $\Theta(2^{n/4})$.
This observation might perhaps be interpreted as evidence of stronger 
expressive power of disjunctive logic programs. A formal definition 
of the appropriate notion of expressiveness and its properties are open
areas of research.

To make the paper self-contained we will now recall the definitions of
a stable model and an answer set \cite{gl88,gl90a}. Let $P$ be a
(disjunctive) propositional logic program built of atoms in the set
$\at$. Let $M\subseteq \at$. By the {\em Gelfond-Lifschitz reduct
of $P$ with respect to $M$}, denoted by $P^M$, we mean the program
obtained from $P$ by:
\begin{enumerate}
\item removing from $P$ all rules with a literal $\n(a)$ in the body,
for some $a\in M$
\item removing all negative literals from all other rules in $P$.
\end{enumerate}

If $P$ is a normal logic program (no disjunctions), $P^M$ is a Horn
program. Consequently, this logic program has its least
model $LM(P^M)$. A set of atoms $M$ is a {\em stable model} of $P$ if
$M=LM(P^M)$.

If $P$ is a disjunctive logic program, instead of the notion of a least
model of $P^M$ (which may not exist), we will use the concept of a minimal 
model. A set of atoms $M$ is an {\em answer set} for $P$ if $M$ is a
minimal model for $P^M$.

The paper is organized as follows. In the next section, we present our
main results on normal logic programs. In particular, we determine
$s(n)$ and characterize the class of extremal logic programs. The
following section discusses the implications of these results for
the design and analysis of algorithms to compute stable models.
In Section \ref{disj}, we study disjunctive logic programs and 
the last section contains conclusions. 
 
\section{NORMAL LOGIC PROGRAMS}
\label{norm}

In this section we study extremal problems for normal
(non-disjunctive) logic programs. We will determine the value of the
function $s(n)$ and we will provide a characterization of all programs
in the class ${\cal LP}_n$ which have $s(n)$ stable models. No bounds on
the length of a clause are needed in this case. It is well known that
each stable model of a program $P$ is a subset of the set of heads of 
$P$. Consequently, $s(n) \leq 2^n$. This bound can easily  be improved.
Stable models of a program form an antichain. Since the size of the
largest antichain in the algebra of subsets of an $n$-element set is
${n \choose {\lfloor n/2 \rfloor}} \approx
0.8\times 2^n/\sqrt{n}$, it clearly follows that,
$s(n) \leq 0.8\times 2^n/\sqrt{n}$.
We will still improve on this bound by showing that $s(n) =
\Theta(3^{n/3}) \approx 
\Theta(2^{0.538n}) << 0.8\times 2^n/\sqrt{n}$. We obtain
similar results for the class ${\cal LP}^2_n$ of logic programs with 
$n$ clauses each of which has at most one literal in the body, and for 
the class ${\cal LP}'_n$ of all logic programs with at most $n$ 
atom occurrences.

Our approach is based on the following version of the notion of reduct
first described in \cite{dix94b} and, independently, in \cite{vsnv95}. 
Let $P$ be a logic program and let $T$ and $F$ be two sets of atoms
such that $T\cap F=\emptyset$. By $simp(P,T,F)$ 
we mean a logic program
obtained from $P$ by
\begin{enumerate}
\item removing all clauses with the head in $T\cup F$
\item removing all clauses that contain an atom from $F$ in the body
\item removing all clauses that contain literal $\n(a)$, where $a\in T$,
in the body
\item removing all atoms $a$, $a\in T$ and literals $\n(a)$, $a\in F$,
from the bodies of all remaining rules.
\end{enumerate}
The simplified program contains all information necessary to reconstruct
stable models of $P$ that contain all atoms from $T$ (``make them
true'') and that do not contain any atoms from $F$ (``make them
false''). The following result was obtained in \cite{dix94b} (see also 
\cite{vsnv95}). We
provide its proof due to the key role this result plays in our considerations.

\begin{lemma}\label{key}
Let $P$ be a logic program and let $T$ and $F$ be disjoint sets of
atoms. If $M$ is a stable model of $P$ such that $T\subseteq M$ and
$M\cap F=\emptyset$, then $M\setminus T$ is a stable model
of $simp(P,T,F)$.
\end{lemma}
\proof Let us define a partition of $P$ into five disjoint programs
$P_1,\ldots,P_5$ (some of them may be empty):
\begin{enumerate}
\item $P_1$ consists of all clauses in $P$ with the head in $T$
\item $P_2$ consists of all clauses in $P$ with the head in $F$
\item $P_3$ consists of all the remaining clauses in $P$ that have 
an atom $a$, where $a\in F$ in the body
\item $P_4$ consists of all the remaining clauses in $P$ that have 
a literal $\n(a)$, where $a\in T$ in the body 
\item $P_5$ consists of all remaining clauses in $P$
\end{enumerate}
It is clear that $simp(P,T,F) = simp(P_5,T,F)$.

Let $M$ be a stable model for $P$ such that $T\subseteq M$ and
$M\cap F=\emptyset$. Since $M$ is the least model of $P^M$, $M$ is 
a model of $P_5^M$. Define $M'=M\setminus T$. We will show that
$M'$ is a model of $simp(P_5,T,F)^M$. Consider a clause
\[
a\lar b_1,\ldots,b_k
\]
from $simp(P_5,T,F)^M$ such that $\{b_1,\ldots,b_k\}\subseteq
M'$. By the definition of Gelfond-Lifschitz reduct, there is a clause
\[
a\lar b_1,\ldots,b_k,\n(c_1)\ldots,\n(c_r)
\]
in $simp(P_5,T,F)$ such that $c_i\notin M$, $1\leq i\leq r$.
Furthermore, by the definition of $simp(P_5,T,F)$, there is a clause
\[
a\lar b_1,\ldots,b_k,b_{k+1},\ldots,b_l,\n(c_1)\ldots,\n(c_r),\n(c_{r+1}),
\ldots,\n(c_s)
\]
in $P_5$ such that $b_i\in T$, $k+1\leq i\leq l$, and $c_i\in F$,
$r+1\leq i\leq s$. Since $F\cap M =\emptyset$, it follows that the clause
\[
a\lar b_1,\ldots,b_k,b_{k+1},\ldots,b_l
\]
belongs to $P_5^M$. Moreover, since $T\subseteq M$, $\{b_1,\ldots,b_l\}
\subseteq M$. Since $M$ is a model of $P_5^M$, $a\in M$. By the
definition of programs $P_i$, $a\notin T$. Hence, $a\in M'$ and,
consequently, $M'$ is a model of $simp(P_5,T,F)^M$.

Consider a model $M''$ of $simp(P_5,T,F)^M$. Assume that $M''\subseteq
M'$. Observe that $M''\cup T$ is a model of $P_1^M$. Since $F\cap
(M''\cup T)=\emptyset$, $M''\cup T$ is a model of $P_3^M$. It is also
clear ($T\subseteq M$) that $P_4^M = \emptyset$. 

Consider a rule
\[
a\lar b_1,\ldots,b_k
\] 
from $P_2^M$. Since $M$ is a model of $P_2^M$ and since $a\notin M$
(recall that $a\in F$ and $M\cap F=\emptyset$), there is $i$, $1\leq
i\leq k$, such that $b_i\notin M$. Since $M''\cup T\subseteq M$, $b_i
\notin M''\cup T$. Thus, any rule in $P_2^M$ is satisfied by $M''\cup
T$.

Finally, consider a rule
\[
a\lar b_1,\ldots,b_l
\]
from $P_5^M$. Assume that $\{b_1,\ldots,b_l\}\subseteq M''\cup T$.
Without loss of generality, we may assume that $\{b_{k+1},\ldots,b_l\}$
are the only $b_i$s that belong to $T$. Then,
$\{b_1,\ldots,b_k\}\subseteq M''$ and 
\[
a\lar b_1,\ldots,b_k
\]
is in $simp(P_5,T,F)^M$. Since $M''$ is a model of $simp(P_5,T,F)^M$,
$a\in M''$. 

Thus, it follows that $M''\cup T$ is a model of $P_5^M$ and,
taking into account the observations made earlier, also of $P^M$.
Since $M''\cup T\subseteq M$ and since $M$ is the least model of $P^M$,
it follows that $M''\cup T=M$. Since $M''\cap T=\emptyset$, it follows
that $M''=M'$. Consequently, $M'$ is the least model of
$simp(P_5,T,F)^M$. By the definition of $P_i$s, it follows that 
$simp(P_5,T,F)^M = simp(P_5,T,F)^{M'}$. Moreover, since 
$simp(P,T,F) = simp(P_5,T,F)$, we have that $simp(P_5,T,F)^M =
simp(P,T,F)^{M'}$. Therefore, $M'$ is the least model of 
$simp(P,T,F)^{M'}$ and, consequently, a stable model of $simp(P,T,F)$.
\halmos

In general, the implication in this result cannot be reversed. However,
it is well known \cite{vsnv95} that if $T$ and $F$ are the sets of atoms 
respectively 
true and false under the well-founded semantics for $P$, then the
converse result holds, too. That is, for every stable model $M'$
of $simp(P,T,F)$, $M'\cup T$ is a stable model of $P$.

Let $P$ be a propositional logic program and let $q$ be an atom.
We define
\begin{enumerate}
\item $P(q^+) = simp(P,\{q\},\emptyset)$
\item $P(q^-)= simp(P,\emptyset,\{q\})$.
\end{enumerate}
Programs $P(q^+)$ and $P(q^-)$ are referred to
as {\em positive} and {\em negative reducts of $P$ with respect to $q$},
respectively. Intuitively, $P(q^+)$ and $P(q^-)$ are 
the programs implied by $P$ and sufficient to determine all stable models 
of $P$. Those stable models of $P$ that contain $q$ can be determined 
from $P(q^+)$, and those stable models of $P$ that do not contain 
$q$, from $P(q^-)$. Formally, we have the following result.

\begin{corollary}\label{cor.a}
Let $P$ be a logic program and $q$ be an atom in $P$. 
\begin{enumerate}
\item Let $M$ be a stable model of $P$. If $q\in M$ then $M \setminus \{q\}$ 
is a stable model of $P(q^+)$. If $q \not\in M$ then $M$ is a stable model 
of $P(q^-)$.
\item $s(P) \leq s(P(q^+)) + s(P(q^-))$.
\end{enumerate}
\end{corollary}

Similarly, we will define now {\em positive} and {\em negative reducts of 
$P$ with respect to a clause $r$}. Assume that  $r = q \lar
a_1,\ldots,a_k, \n(b_1),\ldots,\n(b_l)$. Then, define
\begin{enumerate}
\item $P(r^+) = simp(P,\{q,a_1,\ldots,a_k\},\{b_1,\ldots,b_l\})$, and
\item $P(r^-) = P\setminus \{r\}$. 
\end{enumerate}

We say that a logic program clause $r$ is {\em generating} for a
set of atoms $S$ if every atom occurring positively in the body of $r$
is in $S$ and every atom occurring negated in $r$ is not in $S$. Using
the concept of a generating clause, the intuition behind the definitions
of $P(r^+)$ and $P(r^-)$ is as follows. 
The reduct $P(r^+)$ allows us to compute all those stable models of $P$
for which $r$ {\em is} a generating clause. The reduct $P(r^-)$, on the
other hand, allows us to compute all those stable models of $P$ for which 
$r$ is {\em not} generating. More formally, we have the following lemma.

\begin{corollary}\label{cor.r}
Let $P$ be a logic program 
and $r = q \lar a_1,\ldots,a_k, \n(b_1),\ldots, \n(b_l)$ be a clause of $P$. 
\begin{enumerate}
\item Let $M$ be a stable model of $P$. If $\{a_1,\ldots,a_k\} \subseteq M$ 
and $\{b_1,\ldots,b_l\} \cap M = \emptyset$ then 
$M \setminus \{q,a_1,\ldots,a_k\}$ is a stable model of $P(r^+)$.
Otherwise $M$ is a stable model of $P(r^-)$.
\item $s(P) \leq s(P(r^+)) + s(P(r^-))$. 
\end{enumerate}
\end{corollary}

Also in the case of this result, the implication in its statement cannot
be replaced by equivalence. That is, not every stable model of the reduct 
($P(r^+)$ or $P(r^-)$) gives rise to a stable model of $P$.

It should be clear that Corollaries \ref{cor.a} and \ref{cor.r} imply 
recursive algorithms to compute stable models of a logic program. We will 
discuss these algorithms in the next section. In the remainder of this
section, we will investigate the problem of the maximum number of stable
models of logic programs in classes ${\cal LP}_n$, ${\cal LP}^2_n$
and ${\cal LP}'_n$.

To this end, we will introduce the class of canonical logic programs 
and determine 
for them the number of their stable models . We will use canonical
programs to characterize extremal logic programs in the class ${\cal LP}_n$.

\begin{definition}\label{def.can}
Let $A = \{a_1, a_2, \ldots, a_k\}$ be a set of atoms.
By $c(a_i)$ we denote the clause
\[
c(a_i) = a_i \lar \n(a_1),\ldots,\n(a_{i-1}),\n(a_{i+1}),\ldots,\n(a_k).
\]
A {\em canonical} logic program over $A$, denoted by $CP[A]$,
is the logic program containing exactly $k$ clauses $c(a_1), \ldots, 
c(a_k)$, that is
\[
CP[A] = \bigcup_{i=1}^k \{c(a_i)\}.
\]
\end{definition}

Intuitively, the program $CP[A]$ ``works'' by selecting exactly one atom 
from $A$. Formally, $CP[A]$ has exactly $k$ stable models of the form
$M_i = \{a_i\}$, for $i = 1,\ldots,k$.

\begin{definition}\label{def.234}
Let $P$ be a logic program and $A$ be the set of atoms which appear in $P$. 
Program $P$ is a {\em $2,3,4$-program} if $A$ can be partitioned into 
pairwise disjoint sets $A_1,\ldots,A_l$ such that $2\leq |A_i|\leq 4$
for $i=1,\ldots, l$, and
\[
P = \bigcup_{i=1}^l CP[A_i].
\]
\end{definition}

Roughly speaking, a $2,3,4$-program is a program which arises as a union
of independent canonical programs of sizes 2, 3 or 4. 
A $2,3,4$-program is stratified in the sense of \cite{cho95} and the 
canonical programs are its strata. Stable models of a $2,3,4$-program
can be obtained by selecting (arbitrarily) stable models for each
stratum independently and, then, forming their unions. 

By the {\em
signature} of a $2,3,4$-program $P$ we mean the triple $\la\lambda_2,
\lambda_3,\lambda_4\ra$, where $\lambda_i$, $i=2,3,4$, is the number 
of canonical programs over an $i$-element set appearing in $P$. 

Up to isomorphism, a $2,3,4$-program is uniquely determined by its 
signature. Other basic properties of $2,3,4$-programs are gathered in 
the following proposition (its proof is straightforward and is
omitted). 

\begin{proposition}\label{lem.234}
Let $P$ be a $2,3,4$-program with $n$ clauses and with the signature
$\la\lambda_2,\lambda_3,\lambda_4\ra$. Then:
\begin{enumerate}
\item $n = 2\lambda_2 + 3\lambda_3 +4\lambda_4$,
\item $s(P) = 2^{\lambda_2}  3^{\lambda_3} 4^{\lambda_4}$.
\end{enumerate}
\end{proposition}

As a direct corollary to Proposition \ref{lem.234}, we obtain a result
describing $2,3,4$-programs with $n$ clauses and maximum possible 
number of stable models. For $k\geq 1$, let us define $A(k)$ to be the
unique (up to isomorphism) $2,3,4$-program with the signature
$\la 0,k,0\ra$, and $C(k)$ and $C'(k)$ to be the
unique (up to isomorphism) $2,3,4$-programs with the signatures
$\la 2,k-1,0\ra$ and $\la 0,k-1,1\ra$, respectively. Finally, for $k\geq
0$, let us define $B(k)$ to be the unique (up to isomorphism) 
$2,3,4$-program with the signature $\la 1,k,0\ra$.

\begin{corollary}\label{cor-4}
Let $P$ be a $2,3,4$-program with $n$ clauses and maximum number of 
stable models. Then, 
\begin{enumerate}
\item if $n = 3k$ for some $k\geq 1$, $P=A(k)$,
\item if $n = 3k+1$ for some $k\geq 1$, $P= C(k)$ or $C'(k)$,
\item if $n = 3k+2$ for some $k\geq 0$, $P= B(k)$.
\end{enumerate}
Consequently, the maximum number of stable models of an $2,3,4$-programs 
with $n$ clauses is given by
\[ s_0(n) = \left\{ \begin{array}{ll}
  3 * 3^{\lfloor n/3 \rfloor - 1} & \mbox{for}\ \ n \equiv 0\ mod\ 3\\
  4 * 3^{\lfloor n/3 \rfloor - 1} & \mbox{for}\ \ n \equiv 1\ mod\ 3\\
  6 * 3^{\lfloor n/3 \rfloor - 1} & \mbox{for}\ \ n \equiv 2\ mod\ 3
                      \end{array}
                \right. \]
\end{corollary}

Corollary \ref{cor-4} implies that $s_0(n) = \Theta(3^{n/3})$ and that
\begin{equation} \label{eq-6}
s(n) \geq s_0(n) \geq 3^{n/3}
\end{equation}
We will show that $s(n) = s_0(n)$. We will also
determine the class of all extremal programs. 

We call an atom $q$ occurring in $P$ {\em redundant} if $q$ is 
not the head of a clause in $P$. Let $P$ be a logic program. By
${\overline P}$ we denote the logic
program obtained from $P$ by removing all negated occurrences of
redundant atoms. We define the class ${\cal E}_n$ to consist
of all programs $P$ such that
\begin{enumerate}
\item ${\overline P}$ is $A(k)$, if $n = 3k$ ($k\geq 1$),
\item ${\overline P}$ is $B(k)$, if $n = 3k+2$ ($k\geq 0$), or
\item ${\overline P}$ is $C(k)$ or $C'(k)$, if $n = 3k+1$ ($k\geq 1$).
\end{enumerate} 

\begin{theorem}\label{thm.main}
If $P$ is an extremal logic program with $n \geq 2$ clauses,
then $P$ has $s_0(n)$ stable models. That is, for any $n \geq 2$
\[
s(n) = s_0(n).
\]
In addition, the extremal programs in ${\cal LP}_n$
are exactly the programs in ${\cal E}_n$.
\end{theorem}

Theorem \ref{thm.main} can be proved by induction on $n$. 
The proof relies on Corollaries \ref{cor.a} and \ref{cor.r} that 
establish recursive dependencies between the number of stable 
models of $P$ and of its reducts. It is rather lengthy and, therefore,
we provide it in the appendix. 

The general bound of Theorem \ref{thm.main} can still be slightly
improved (lowered) if the class of programs is further restricted.
Since there are extremal programs for the whole class ${\cal LP}_n$ with
no more than 2 literals in the body of each clause, the only reasonable
restriction is to limit the number of literal occurrences in the body to
at most 1. The class of programs with $n$ clauses and satisfying this
restriction will be denoted by ${\cal LP}^2_n$. 

Denote by $P(k)$ a $2,3,4$-program with signature $\la k,0,0\ra$. 
Clearly, $P(k) \in {\cal LP}^2_n$. We have the following result. The
proof uses similar techniques as the proof of Theorem \ref{thm.main}
and is omitted.

\begin{theorem}\label{body-1}
For every program $P\in {\cal LP}^2_n$, $s(P) \leq 2^{\lfloor n/2\rfloor}$.
Moreover, there are programs in ${\cal LP}^2_n$ for which this bound is 
attained. Program $P(k)$ is a unique (up to isomorphism) extremal
program with $n=2k$ clauses, and every extremal program with $n=2k+1$ 
clauses can
be obtained by adding one more  clause to $P(k)$ of one of the following
forms: $p\lar a$, $a\lar$, and $a\lar \n(b)$, where $p$ is an arbitrary
atom (may or may not occur in $P(k)$), and $a$ and $b$ are atoms not
occurring in $P(k)$.
\end{theorem}

Next, we will consider the class ${\cal LP}'_n$ of all logic programs with 
the total size (number of literal occurrences in the bodies and heads) at 
most $n$. Let $s'(n)$ be defined as the maximum number of stable models
for a program in ${\cal LP}'_n$. We have the following result.

\begin{theorem}\label{total-n}
For every integer $n\geq 1$, $s'(n) = \Theta(2^{n/4})$.
\end{theorem}
\proof We will show that for every $n\geq 1$, and for every logic
program of size at most $n$, $s(P)\leq 2^{n/4}$. We will proceed by
induction. Consider a logic program $P$ such that the size of $P$
is at most $4$. If $P$ has one rule, then it has at most one stable
model. If $P$ has two rules and one of them is a fact (rule with empty 
body), then $P$ has at most one stable model. Otherwise, 
$P\in {\cal LP}^2_n$ and $s(P)\leq 2^{n/4}$ follows from 
Theorem \ref{body-1}. If $P$ has 
three rules, then at least two of these rules are facts and $P$
has at most one stable model. If $P$ has four rules, it is a Horn
program and has exactly one stable model. Hence, in all these cases,
$s(P)\leq 2^{n/4}$. Since $P$ has size 4, it has at most four rules and
the basis of induction is established.

Consider now a logic program $P$ of size $n > 4$. Assume that $P$
has a rule, $r$, with at least two elements in its body. Let $a$ be 
the head
of $r$. If $a$ and $\n(a)$ do not occur in the body of any rule in
$P\setminus\{r\}$, then $s(P)\leq s(P\setminus\{r\})$ and the result
follows by the induction hypothesis. So, assume that there is a rule in
$P\setminus\{r\}$ such that $a$ or $\n(a)$ occurs in its body. Then,
both $P(a^+)$ and $P(a^-)$ have sizes at most $n-4$.
By Corollary \ref{cor.a}, $s(P)\leq s(P(a^+))+s(P(a^-))$. Consequently,
by  the induction hypothesis, $s(P) \leq 2^{n/4}$.

Thus, assume that each rule in $P$ has at most one literal in its body.
If at least one of these rules, say $r$, has empty body, 
then every stable model of $P$ contains the head of $r$ (say $a$). Thus,
$s(P)\leq P(a^+)$ (Corollary \ref{cor.a}) and the result follows by 
the induction hypothesis.

Hence, assume that each rule in $P$ has nonempty body.
Let $p$ be the number of rules in $P$. Then, $p\leq \lfloor n/2\rfloor$. 
Moreover, $P\in {\cal LP}_p^2$. By Theorem \ref{body-1},
$s(P) \leq 2^{\lfloor p/2\rfloor}  \leq 2^{n/4}$.
\halmos 

Finally, let us observe that every antichain $\cal F$ of sets of atoms
is representable by a logic program.

\begin{theorem}\label{rep}
For every antichain $\cal F$ of finite sets there is a logic program $P$
such that $\St(P) = {\cal F}$. Moreover, there exists such $P$ with at
most $\sum_{B\in {\cal F}} |B|$ clauses and total size at most $|{\cal
F}|\times \sum_{B\in {\cal F}} |B|$.
\end{theorem}
\proof Consider a finite antichain $\cal F$ of finite sets.
Let $B\in {\cal F}$.  For every $C\in {\cal F}$, $B\not= C$,
denote by $x_{B,C}$ an element from $C\setminus B$ (it is possible
as $\cal F$ is an antichain). Now, for each element $b\in B$, define
\[
r_b = \ \ \ b\lar \n(x_{B,C_1}),\ldots,\n(x_{B,C_k}),
\]
where $C_1,\ldots,C_k$ are all elements of $\cal F$ {\em other} than
$B$. Next, define a program $P_B$ to consist of all rules $r_b$, for
$b\in B$. Finally, define
\[
P_{\cal F} = \bigcup_{B\in{\cal F}} P_B.
\]
It is easy to verify that $\St(P_{\cal F}) = {\cal F}$ and that the size
of $P_{\cal F}$ is $|{\cal F}|\times \sum_{B\in {\cal F}} |B|$.
\halmos

On one hand this theorem states that logic programs can encode any
antichain $\cal F$. On the other, the encoding that is guaranteed by 
this result is quite large (in fact, larger than the explicit encoding of 
$\cal F$). In the same time, our earlier results show that often
substantial compression can be achieved. In particular, there are
antichains of the total size of $\Theta(n3^{n/3})$ that can be encoded
by logic programs of size $\Theta(n)$. More in-depth understanding
of applicability of logic programming as a tool to concisely
represent antichains of sets remains an open area of investigation.

\section{APPLICATIONS IN STABLE MODEL COMPUTATION}
\label{algs}

In this section we will describe algorithms for computing
stable models of logic programs. These algorithms are recursive 
and are implied by Corollaries \ref{cor.a} and \ref{cor.r}. 
They select an atom (or a clause, in the case of Corollary \ref{cor.r})
and compute the corresponding reducts. According to Corollaries \ref{cor.a}
and \ref{cor.r}, stable models of $P$ can be reconstructed from stable
models of the reducts. However, it is not, in general, the case that every
stable model of a reduct implies a stable model of $P$ (see the comments
after Corollary \ref{cor.r}). Therefore, all candidates for stable models 
for $P$, that are produced out of the stable models of the reduct, must 
be tested for stability for $P$. To this end, an auxiliary procedure 
{\sc is\_stable} is used. Calling {\sc is\_stable} for a set of atoms
$M$ and a logic program $P$ returns {\bf true} if $M$ is a
stable model of $P$, and it returns {\bf false}, otherwise. 

In our algorithms we use yet another auxiliary procedure, 
{\sc implied\_set}.
This procedure takes one input parameter, a logic program $P$, and
outputs a set of atoms $M$ and a logic program $P_0$ (modified $P$)
with the following properties:
\begin{enumerate}
\item $M$ is a subset of every stable model of $P$, and
\item stable models of $P$ are exactly the unions of $M$ 
and stable models of $P_0$.
\end{enumerate}
There are several specific choices for the procedure {\sc implied\_set}.
A trivial option is to return $M = \emptyset$ and $P_0 = P$. Another
possibility is implied by our comments following the proof of 
Lemma \ref{key}. Let $T$ and $F$ be sets of
atoms that are true and false, respectively, under the well-founded
semantics for $P$. The procedure {\sc implied\_set} might return $T$ as $M$,
the program $simp(P,T,F)$ as $P_0$. This choice turned out to be
critical to the performance of the s-models system \cite{ns96} and, we
expect, it will lead to significant speedups once our algorithms are
implemented. However, in general, there are many other,
intermediate, ways to compute $M$ and $P_0$ in polynomial time
so that conditions (1) and (2) above are satisfied. Experimental
studies are necessary to compare these defferent choices among each
other (this is a subject of an ongoing work).

\begin{figure}[t]
\rule{2.00in}{0.5mm}\\
\noindent
{\sc stable\_models\_a$(P)$}\\
{\it Input:} a finite logic program $P$;\\
{\it Returns:} family $Q$ of all stable models of $P$;\\
\ \\
{\sc implied\_set}($P,M,P_0$);\\
{\bf if} $(|P_0| = 0)$ {\bf then} {\bf return} $\{M\}$\\
{\bf else}\\
\hspace*{0.3in}$Q := \emptyset$;\\
\hspace*{0.3in} $q := $ {\sc select\_atom}($P_0$);\\
\ \\
\hspace*{0.3in} $P_1 := P_0(q^+)$;\\
\hspace*{0.3in} $L := $ {\sc stable\_models\_a}($P_1$);\\
\hspace*{0.3in} {\bf for all} $N \in L$ {\bf do}
{\bf if} {\sc is\_stable}($P_0,\{q\}\cup N$) {\bf then} 
		$Q := Q \cup \{M \cup \{q\} \cup N\}$;\\
\ \\
\hspace*{0.3in} $P_2 := P_0(q^-)$;\\
\hspace*{0.3in} $L := $ {\sc stable\_models\_a}($P_2$);\\
\hspace*{0.3in} {\bf for all} $N \in L$ {\bf do}
{\bf if} {\sc is\_stable}($P_0,N$) {\bf then} 
                $ Q := Q \cup \{M \cup N\}$;\\
\ \\
\hspace*{0.3in}{\bf return} $Q$;\\
\rule{2.00in}{0.5mm}
\caption{Algorithm for computing stable models by splitting on atoms.}
\label{fig.alg1}
\end{figure}

We will now describe the algorithms. We adopt the following notation. 
For a logic program clause $r$, by $\hd(r)$ we denote the head of $r$ 
and by $\pb(r)$, the set of atoms occurring positively in the body of $r$.

First, we will discuss an algorithm based on splitting the original program
(that is, computing the reducts) with respect to a selected atom. This
idea and the resulting algorithm appeared first in \cite{vsnv95}. The
correctness of this method is guaranteed by Lemma \ref{key} (or, 
more specifically, by Corollary \ref{cor.a}). We call 
this algorithm {\sc stable\_models\_a}. 

In this algorithm, to compute stable models
for an input program $P$ we first simplify it to a program $P_0$
by executing the
procedure {\sc implied\_set}. A set of atoms $M$ contained in all stable
models of $P$ is also computed. Due to our requirements on the {\sc
implied\_set} procedure, at this point, to compute all models of
$P$, we need to compute all models of $P_0$ and expand each by $M$.
To this end, we select an atom occurring in $P_0$, say $q$, 
by calling a procedure {\sc select\_atom}. Then, we compute 
the reducts $P_0(q^+)$ and  $P_0(q^-)$. For both reducts we compute
their stable models. Each of these stable models gives rise to a set of
atoms $\{q\}\cup N$ (in the case of stable models for $P_0(q^+)$)
or $N$ (in the case of stable models for $P_0(q^-)$). Each of these
sets is a candidate for a stable model for $P_0$. Calls to the
procedure {\sc is\_stable} determine those that are. These sets,
expanded by $M$, are returned as the stable models of $P$.
We present the pseudocode
for this algorithm in Figure \ref{fig.alg1}.

\begin{figure}[t]
\rule{2.00in}{0.5mm}\\
\noindent
{\sc stable\_models\_r$(P)$}\\
{\it Input:} a finite logic program $P$;\\
{\it Returns:} family $Q$ of all stable models of $P$;\\
\ \\
{\sc implied\_set}($P,M,P_0$);\\
{\bf if} $(|P_0| = 0)$ {\bf then} {\bf return} $\{M\}$\\
{\bf else}\\
\hspace*{0.3in}$Q := \emptyset$;\\
\hspace*{0.3in} $r := $ {\sc select\_clause}($P_0$);\\
\ \\
\hspace*{0.3in} $P_1 := P_0(r^+)$;\\
\hspace*{0.3in} $L := $\  {\sc stable\_models\_r}($P_1$);\\
\hspace*{0.3in} {\bf for all} $N \in L$ {\bf do}
 {\bf if}\  {\sc is\_stable}($P_0, N \cup \pb(r)\cup\{\hd(r)\}$) \\
\hspace*{0.6in} {\bf then}
                $ Q := Q \cup \{M \cup N \cup \pb(r)\cup\{\hd(r)\}\}$;\\
\ \\
\hspace*{0.3in} $P_2 := P_0(r^-)$;\\
\hspace*{0.3in} $L := $\ {\sc stable\_models\_r}($P_2$);\\
\hspace*{0.3in} {\bf for all} $N \in L$ {\bf do}
 {\bf if}\  {\sc is\_stable}($P_0,N$) {\bf then}
                $ Q := Q \cup \{M \cup N\}$;\\
\ \\
\hspace*{0.3in} {\bf return} $Q$;\\
\rule{2.00in}{0.5mm}
\caption{Algorithm for computing stable models by splitting on clauses.}
\label{fig.alg2}
\end{figure}

The second algorithm, {\sc stable\_models\_r}, is similar. It
is based on Corollary \ref{cor.r}. That is, instead of trying to find
stable models of $P$ among the sets of atoms implied by the stable 
models of $P(q^+)$ and  $P(q^-)$, we search for stable models of $P$ 
using stable models of $P(r^+)$ and $P(r^-)$, where $r$ is a {\em clause}
of $P$. The correctness of this approach follows by Corollary \ref{cor.r}. 
The pseudocode is given in Figure \ref{fig.alg2}.

Algorithms {\sc stable\_models\_a} and {\sc stable\_models\_r}
can easily be merged together into a hybrid method, which we call 
{\sc stable\_models\_h} (Figure \ref{fig.hyb}). Here, in each recursive
call to {\sc stable\_models\_h} we start by deciding whether the
splitting (reduct computation) will be performed with respect to an
atom or to a clause. 
The function {\sc select\_mode}(``atom'',``clause'')
makes this decision.
Then, depending on the outcome, the algorithm follows 
the approach of either {\sc stable\_models\_a} or {\sc stable\_models\_r}.
That is, either an atom or a clause is selected, the corresponding reducts 
are computed and recursive calls to {\sc stable\_models\_h} are made.

All three algorithms provide a convenient framework for experimentation
with different heuristics for pruning the search space of all subsets
of the set of atoms. In general, the performance of these algorithms
depends heavily on how the selection routines {\sc select\_atom},
{\sc select\_clause} and {\sc select\_mode} are implemented.
Although any selection strategy yields a correct algorithm, some
approaches are more efficient than others.
In particular, the proof of Theorem \ref{thm.main} implies selecting
techniques for the algorithm {\sc stable\_models\_h} guaranteeing
that the algorithm terminates after the total of at most $O(3^{n/3})$
recursive calls.

Let us also observe that the recursive dependencies given in
Corollaries \ref{cor.a} and \ref{cor.r} indicate that in order to keep
the search space (number of recursive calls) small, selection heuristics
should attempt to keep the total size of $P(q^+) \cup P(q^-)$
or $P(r^+) \cup P(r^-)$ as small as possible.

{\small
\begin{figure}[h]
\rule{2.00in}{0.5mm}\\
\noindent
{\sc stable\_models\_h$(P)$}\\
{\it Input:} a finite logic program $P$;\\
{\it Returns:} family $Q$ of all stable models of $P$;\\
\ \\
{\sc implied\_set}($P,M,P_0$);\\
{\bf if} $(|P_0| = 0)$ {\bf then} {\bf return} $\{M\}$\\
{\bf else}\\
\hspace*{0.3in}$Q := \emptyset$;\\
\hspace*{0.3in}$split\_mode := $ {\sc select\_mode}(``atom'',``clause'');\\

\hspace*{0.3in}{\bf if} $(split\_mode = \mbox{``atom''})$\ {\bf then}\\
\hspace*{0.45in} {\bf begin}\\
\hspace*{0.6in} $q := $ {\sc select\_atom}($P_0$);\\
\hspace*{0.6in} $P_1 := P_0(q^+)$;\\
\hspace*{0.6in} $L := $ {\sc stable\_models\_h}($P_1$);\\
\hspace*{0.6in} {\bf for all} $N \in L$ {\bf do}
{\bf if} {\sc is\_stable}($P_0,\{q\}\cup N$) {\bf then}
                $Q := Q \cup \{M \cup \{q\} \cup N\}$;\\
\hspace*{0.6in} $P_2 := P_0(q^-)$;\\
\hspace*{0.6in} $L := $ {\sc stable\_models\_h}($P_2$);\\
\hspace*{0.6in} {\bf for all} $N \in L$ {\bf do}
{\bf if} {\sc is\_stable}($P_0,N$) {\bf then}
                $ Q := Q \cup \{M \cup N\}$;\\
\hspace*{0.45in} {\bf end}\\
\hspace*{0.3in}{\bf else} \hspace*{0.2in} $(\ast\  
                     split\_mode = \mbox{``clause''}\  \ast)$\\
\hspace*{0.45in} {\bf begin}\\
\hspace*{0.6in} $r := $ {\sc select\_clause}($P_0$);\\
\hspace*{0.6in} $P_1 := P_0(r^+)$;\\
\hspace*{0.6in} $L := $\  {\sc stable\_models\_h}($P_1$);\\
\hspace*{0.6in} {\bf for all} $N \in L$ {\bf do}
 {\bf if}\  {\sc is\_stable}($P_0, N \cup \pb(r)\cup\{\hd(r)\}$)\\
\hspace*{0.75in} {\bf then}
                $ Q := Q \cup \{M \cup N \cup \pb(r)\cup\{\hd(r)\}\}$;\\
\hspace*{0.6in} $P_2 := P_0(r^-)$;\\
\hspace*{0.6in} $L := $\ {\sc stable\_models\_h}($P_2$);\\
\hspace*{0.6in} {\bf for all} $N \in L$ {\bf do}
 {\bf if}\  {\sc is\_stable}($P_0,N$) {\bf then}
                $ Q := Q \cup \{M \cup N\}$;\\
\hspace*{0.45in} {\bf end}\\
\hspace*{0.3in}{\bf return} $Q$;\\
\rule{2.00in}{0.5mm}
\caption{Hybrid algorithm for computing stable models.}
\label{fig.hyb}
\end{figure}
}

The presented algorithms compute all stable models for the input program
$P$. They
can be easily modified to handle other tasks associated with
logic programming. That is, they can be tailored to compute one stable
model, determine whether a stable model for $P$ exists, as well as answer
whether an atom is true or false in all stable models of $P$
(cautious reasoning), or in one model of $P$ (brave reasoning). All
these
tasks can be accomplished by adding a suitable stop function and by
halting the algorithm as soon as the query can be answered.

The general structure of our algorithms is similar to well-known
Davis-Putnam method for satisfiability problem. The {\sc implied\_set}
procedure corresponds to the, so called, unit-propagation phase of
Davis-Putnam algorithm. In this phase necessary and easy-to-compute
conclusions of the current state are drawn to reduce the search space.
If the answer is still unknown then a guess is needed and
two recursive calls are performed to try both possibilities.
But there are also differences. First, in our case, splitting can also
be done with respect to a clause. The second difference is due
to nonmonotonicity of stable semantics for logic programs. 
When a recursive call in Davis-Putnam
procedure returns an answer, this answer is guaranteed to be correct.
There is no such guarantee in the case of stable models. Each answer
(stable model) returned by a recursive call in our algorithms must
be additionally tested (by {\sc is\_stable} procedure) to see whether 
it is a stable model for the original program.

\section{DISJUNCTIVE LOGIC PROGRAMS}
\label{disj}

In this section, we will focus on the class of disjunctive logic
programs ${\cal DP}_{n,m}$.
For a set of atoms $\{a_1,\ldots, a_m\}$, let us denote by 
$d(a_1,\ldots,a_m)$ the disjunctive clause of the form
\[
a_1\vee\ldots\vee a_k \lar.
\]
By $D(n,m)$, we will denote the disjunctive logic program consisting of
$n$ clauses:
\[
d(a_{1,1},\ldots,a_{1,m})
\]
\[
\cdots
\]
\[
d(a_{n,1},\ldots,a_{n,m}),
\]
with all atoms $a_{i,j}$ --- distinct. It is clear that every set
of the form 
\[
\{a_{i,j_i}\colon i=1,\ldots, n,\ 1\leq j_i\leq m\}
\]
is an answer set for $D(n,m)$, and that all answer sets for $D(n,m)$ are
of this form. Hence, 
\[
|\St(D(n,m))| = m^n.
\]
Consequently, general upper bounds on the number of answer sets
for disjunctive programs in such classes that allow clauses of
arbitrary length do not exist.

Turning attention to the class ${\cal DP}_{n,m}$, it is now clear that,
since $D(n,m)\in {\cal DP}_{n,m}$, 
\[
d(n,m) \geq m^n.
\]
The main result of this section shows that, in fact,
\[
d(n,m) = m^n
\]
and the program $D(n,m)$ is the only (up to isomorphism)
extremal program in this class.

Consider a clause $d$ of the form
\[
a_1\vee \ldots \vee a_k \lar b_1,\ldots , b_p,\n(c_1),\ldots,\n(c_q).
\]
By $d^+$ we will denote the clause obtained from $d$ by moving all
negated atoms to the head. That is, $d^+$ is of the form:
\[
a_1\vee \ldots \vee a_k\vee c_1 \vee \ldots \vee c_q \lar b_1,\ldots ,
b_p.
\]
Let $D$ be a disjunctive program. Define
\[
D^+ = \{d^+\colon d\in D\}.
\]

\begin{lemma}\label{lem-1}
For every disjunctive logic program $D$, $\St(D)\subseteq \St(D^+)$.
\end{lemma}
\proof Let $M\in \St(D)$. Then, $M$ is a minimal model of
the Gelfond-Lifschitz reduct $D^M$ and, as is well-known, $M$ is a model
of $D$. It follows that $M$ is a model of $D^+$. To show that $M\in
\St(D^+)$, we need to show that $M$ is a minimal model of $D^+$.

Consider a model $M'$ of $D^+$ and assume that $M'\subseteq M$. Take a
clause
\[
a_1\vee \ldots \vee a_k \lar b_1,\ldots , b_m
\]
from $D^M$. Then, there is a rule
\[
a_1\vee \ldots \vee a_k \lar b_1,\ldots , b_m,\n(c_1),\ldots,\n(c_n)
\]
in $D$ such that $n\geq 0$ and $c_1,\ldots, c_n \not\in M$. 
Since $M'\subseteq M$, $c_1,\ldots, c_n \not\in M'$.
Assume that $\{b_1,\ldots,b_m\}\subseteq M'$. Then, since $M'$ is a
model of $D$ (recall that it is a model of $D^+$), there is $i$,
$1\leq i\leq k$, such that $a_i\in M'$.
It follows that $M'$ is a model of $D^M$. Since $M$ is a minimal model of
$D^M$, $M=M'$. Hence, $M$ is a minimal model of $D^+$. \halmos

Lemma \ref{lem-1} allows us to restrict our search for disjunctive
programs with the largest number of answer sets to those programs
that do not contain negated occurrences of atoms. 

\begin{lemma}
\label{aux-1}
Let $D$ be a disjunctive program with $n$ rules $d_1,\ldots, d_n$.
Assume that for each $i$, $1\leq i\leq n$, $d_i$ has empty body and
exactly $h_i$ different disjuncts in the head. Then $D$ has at most
$h_1\times \cdots\times h_n$ answer sets. Moreover, if $D$ has exactly
$h_1\times \cdots\times h_n$ different answer sets, then no two rules
have the same atom in their heads.
\end{lemma}
\proof Clearly, for each program whose every rule has empty body,
answer sets are exactly minimal models. So, we have to prove that
$D$ has at most $h_1\times \cdots\times h_n$ minimal models.
We will proceed by induction on the size of $D$ (total number of literal
occurrences in $D$). If the size of $D$ is 1, the assertion holds.
Consider now a disjunctive logic program $D$ of size $k> 1$, whose 
each rule has empty body. Assume $D$ has $n$ rules $d_1,\ldots, d_n$ and
that for each $i$, $1\leq i\leq n$, $d_i$ has exactly $h_i$ different
disjuncts in the head. 

Consider a minimal model $M$ of $D$. Let $a$ be any atom appearing in
the head of $d_1$. Let $M$ be a minimal model of $D$. Assume that 
$a\notin M$. Then, $M$ is a minimal model of a program $D'$ obtained
from $D$ by removing $a$ from the head of each rule in which it appears.
By induction hypothesis applied to $D'$, there are at most
$(h_1-1)\times h_2\times \cdots\times h_n$ minimal models $M$ of $D$
that do not contain $a$. Moreover, this number 
equals $(h_1-1)\times h_2\times \cdots\times h_n$ precisely if the heads 
of rules of $D'$ have $h_1-1$, 
$h_2, \ldots, h_n$ disjuncts in their heads, and if no atom appears in
$D'$ more than once. This happens precisely when no atom
appears more than once in $D$.

The other possibility for $M$ is that $a\in M$. In this case, define
$D'$ to be a program obtained from $D$ by removing all clauses with $a$
in the head (in particular, $d_1$ is removed). Assume that
$D'=\{d_{i_1},\ldots,d_{i_p}\}$. Since $d_1$ is removed, $p<n$.
Clearly, $M\setminus \{a\}$ is a minimal model of $D'$. If $D'\not=\emptyset$, 
by induction hypothesis, it follows that there are at most $h_{i_1}\times
\cdots\times h_{i_p}\leq h_2\times \cdots\times h_n$ minimal model of $D$ 
that contain $a$. Moreover, this number equals $h_2\times \cdots\times
h_n$ occurs precisely when $a$ occurs only in $d_1$ and if no
atom appears more than once in $d_2,\ldots, d_n$.

It follows that the total number of minimal models of $D$ is at most 
\[
(h_1-1)\times h_2\times \cdots\times h_n + h_2\times \cdots\times h_n =
h_1\times h_2\times \cdots\times h_n.
\]
It also follows that the number of minimal models of $D$ is
$h_1\times \cdots\times h_n$ if and only if no atom appears in $D$ more
than once. \halmos

\begin{theorem}\label{th-1}
For every integers $m\geq 1$ and $n\geq 1$, and for every program $D\in
{\cal DP}_{n,m}$, $|\St(D)|\leq m^n$. Moreover, the program $D(n,m)$ is
the only program in the class ${\cal DP}_{n,m}$ for which the bound of
$m^n$ is reached. In particular, $d(n,m) = m^n$.
\end{theorem}
\proof We will proceed by induction on $n$.
The theorem clearly holds if $n=1$. It is also true if $m=1$.
So, assume that $m\geq 2$ and $n\geq 2$.

We will first focus on disjunctive programs in ${\cal DP}_{n,m}$
that do not contain negated occurrences of atoms. Let $D\in {\cal DP}_{n,m}$ 
be such a program, say $D=\{d_1,\ldots,d_n\}$. Assume that the rule
$d_i$ has $h_i$ atoms in its head. 

If each clause in $D$ has a nonempty body, $D$ has
exactly one answer set model, the empty set. Since $m\geq 2$, $s(D) < m^n$
(the inequality holds and $D$ is not extremal).

Next, assume that at least one rule in $D$ has empty body.
Let $D'$ be a subset of $D$ consisting of all the clauses with the
empty body. Let $n'$ denote the number of clauses in $D'$. Hence,
$n'>0$. Each minimal model for $D$ can be obtained by
the following procedure: 
\begin{enumerate}
\item Pick a minimal model $M'$ of $D'$. If $D =D'$, output $M'$ and
stop.
\item Otherwise, reduce $D\setminus D'$ by removing clauses satisfied by $M'$
as well as atoms from the bodies of the remaining rules that belong to $M'$. 
Call the resulting program $D''$.
\item Pick a minimal model $M''$ of $D''$.
\item Output $M'\cup M''$ as a minimal model of $D$.
\end{enumerate}
Clearly, Lemma \ref{aux-1} applies to $D'$. Hence,
$|\St(D')|\leq m^{n'}$, with equality if and only if $D'=D({n',m})$.
If $D''=\emptyset$, then there is only one possibility for $M''$, namely
$M''=\emptyset$.
If $D''\not=\emptyset$, $D''\in {\cal DP}_{n'',m}$, for some $n''\leq n-n'
<n$.
By induction hypothesis, $|\St(D'')|\leq m^{n''}$. Moreover,
equality holds if and only if $D'' = D({n'',m})$. Consequently,
$|\St(D)|\leq m^{n'}\times  m^{n''} \leq m^n$, with equality holding
if and only if $D = D(n,m)$. 

Consider now an arbitrary program $D\in {\cal DP}_{n,m}$. Assume that
$D$ is extremal. It follows from Lemma \ref{lem-1} that $D^+$ is also
extremal. Hence, $D^+ = D(n,m)$. Assume that $D\not=D^+$. Then, there
is a rule in $D$ that contains at least one negated atom, say $a$. It 
follows from the definitions of $D^+$ and $D(n,m)$, and from the
equality $D^+ = D(n,m)$ that:
\begin{enumerate}
\item there is an answer set $M$ of $D^+$ such that $a\in M$, and 
\item no answer set for $D$ contains $a$.
\end{enumerate}
Since $\St(D)\subseteq \St(D^+)$, and since $D^+$ is extremal, it follows
that $D$ is not extremal, a contradiction. Hence, $D=D^+ =D(n,m)$.
\halmos

Finally, we will consider the class ${\cal DP}_n$ of all logic programs
with the total size (number of literal occurrences in the bodies and heads)
at most $n$. Let $d'(n)$ be defined as the maximum number of answer sets
for a disjunctive program in ${\cal DP}_n$. We have the following result.

\begin{theorem}\label{d-total-n}
For every $n\geq 2$, $d'(n) = \Theta(2^{n/2})$.
\end{theorem}
\proof Assume that $D$ has size $n$ and that it has $k$ rules.
By Theorem \ref{th-1} it follows that $|\St(D)|\leq m^k$, where
$m = \lceil n/k\rceil$. The value $m^k$, under the constraint
$m = \lceil n/k\rceil$, assumes its maximum for $k = \lfloor
n/2\rfloor$. Hence, for every disjunctive logic program $D$ of size $n$,
$|\St(D)| = O(2^{n/2})$. In the same time, program $D({\lfloor
n/2\rfloor,2})$ demonstrates that there is a disjunctive program $D$
of size at most $n$ such that $|\St(D)| = \Omega(2^{n/2})$. Hence, the
assertion follows. \halmos

Compared with the estimate from Theorem \ref{total-n} for the function
$s'(n)$, the function $d'(n)$ is much larger (it is, roughly the square
of $s'(n)$. Consequently, there are antichains representable by
disjunctive logic programs with the cardinality of the order of the
square of the cardinality of largest antichains representable by logic
programs of the same total size. This may be an additional argument for 
disjunctive logic programs as a knowledge representation mechanism.

\section{CONCLUSIONS}
\label{concl}

In this paper, we studied extremal problems appearing in the area of
logic programming. Specifically, we were interested in the maximum
number of stable models (answer sets) a program  (disjunctive program)
from a given class may have. We have studied several classes in detail.
We determined the maximum number of stable models for logic programs
with $n$ clauses. Similarly, this maximum was also established for
logic programs with $n$ clauses, each of length at
most 2, and for logic programs of total size at most $n$. In some of
these cases we also characterized the extremal programs, that is, the
programs for which the maxima are attained. Similar results were
obtained for disjunctive logic programs. Our results have interesting
algorithmic implications. Several algorithms, having a flavor of
Davis-Putnam procedure, for computing stable model 
semantics are presented in the paper. 

Extremal problems for logic programming have not been studied
so far. This paper shows that they deserve more attention. They
are interesting in their own right and have interesting computational
and knowledge representation applications.

\section{APPENDIX -- PROOF OF THE MAIN RESULT}

First, we prove auxiliary lemmas which will be used in the proof of
Theorem \ref{thm.main}

\noindent
\begin{lemma}\label{f.incr}
For any $n \leq 1$, $s(n) < s(n+1)$.
\end{lemma}
\proof Let $P$ be a program with $n$ rules and $s(P)$ stable models.
To complete the proof it is enough to show that there is a logic program
$P'$ with $n+1$ rules and $s(P) < s(P')$.
Assume first that $s(P) \leq 1$. Then, as $P'$ we can take 
any program with $n+1$ rules and 2 or more stable models 
(since $n+1 \geq 2$, such programs exist).

Suppose now, that $P$ has at least 2 stable models. Let $M_1, M_2,...,M_k$ be
the all stable models of $P$. We construct $P'$ as follows. Since stable
models of a logic program form an antichain, 
every model $M_i$, $1 \leq i \leq k$, is not empty.
Let $b$ be a propositional atom not occurring in $P$.
Let $A = \{a_1,a_2,\ldots,a_l\}$ be any set of atoms such that for all
$i, 1 \leq i \leq k$, $A \cap M_i \neq \emptyset$. Finally, let
\begin{eqnarray*}
\lefteqn{P' = \{ head(r) \lar body(r),\n(b): r \in P\} \cup}\\
& & \{b \lar \n(a_1),\n(a_2),\ldots,\n(a_l)\}
\end{eqnarray*}
It is easy to see that
$M_1, M_2,...,M_k, \{b\}$ are stable models for $P'$. Thus, the proof
of the lemma is complete.
\halmos

A clause $r$ of $P$ is called
{\em redundant} if the head of $r$ occurs (negated or not) in the body
of $r$, or if there is an atom $q$ such that both $q$ and $\n(q)$
occur in the body of $r$.

\begin{lemma}\label{bounds.1}
If $P$ is an extremal program with $n \geq 2$ rules than:
\begin{enumerate}
\item $P$ contains no positive redundant literals,
\item $P$ contains no redundant rules,
\item $P$ contains no facts (i.e. rules with empty body),
\item every head of a rule in $P$ appears in the body of another rule in
$P$.
\end{enumerate}
\end{lemma}
\vspace*{0.1in}
\proof If $P$ contains a positive redundant literal $q$ in the body
of a rule $r$ then every stable model for $P$ is a stable model for $P(r^-)$.
Hence $\St(P) \subseteq \St(P(r^-)$. So, from Lemma \ref{f.incr}, we have that
\[
s(P) \leq s(P(r^-)) \leq s(n-1) < s(n).
\]
This means that $P$ is not extremal.

If $P$ contains a redundant rule $r$ then stable
models of $P$ are exactly the stable models
of $P(r^-)$. Again, $P$ is not extremal. If $P$ contains a fact $q \lar$
then $q$ must belong to every stable model of $P$. That is, 
\[
s(P) \leq s(P(q^+)) \leq s(n-1) < s(n),
\]
and $P$ is not extremal.

Assume that $P$ contains a rule $r$ with head $q$ and $q$ does not 
appear negatively or positively in the body of any other rule.
For any set of atoms $M$, $M$ is a stable model for $P$ if and only if 
$M \smn \{q\}$ is a stable model for $P(q^+)$. Hence, again 
$s(P) \leq s(P(q^+))  < s(n)$
and $P$ is not an extremal program. 
\halmos

\begin{lemma}\label{inq.f0}
Let $n$ be a positive integer and $n = 3m + l$, where $0 \leq l \leq 2$.
For any $n \geq 3$
\begin{equation}\label{inq.f0a}
s_0(n) \geq 2s_0(n-2).
\end{equation}
Moreover, if $l=0$ then $s_0(n) > 2s_0(n-2)$, otherwise
$s_0(n) = 2s_0(n-2)$.

\noindent
For any two integers $x,y$, such that $x,y \geq 2$ and $2 < \max(x,y) < n$,
\begin{equation}\label{inq.f0b}
s_0(n) > s_0(n-x) + s_0(n-y).
\end{equation}

\noindent
For any $n \geq 5$
\begin{equation}\label{inq.f0c}
s_0(n) \geq s_0(n-1) + s_0(n-4).
\end{equation}
Moreover, if $l=1$ then $s_0(n) = s_0(n-1) + s_0(n-4)$, otherwise
$s_0(n) > s_0(n-1) + s_0(n-4) $.

\noindent
For any integer $x$, such that $4 < x < n$,
\begin{equation}\label{inq.f0d}
s_0(n) > s_0(n-1) + s_0(n-x).
\end{equation}
\end{lemma}
\proof Straightforward arithmetic for inequalities (\ref{inq.f0a})
and (\ref{inq.f0c}). Inequalities (\ref{inq.f0b}) and (\ref{inq.f0d})
are implied by (\ref{inq.f0a}) and (\ref{inq.f0c}) and mononicity
of $s_0$. \halmos

\begin{lemma} 
\label{aux}
Let $P$ be a logic program with $n$ rules with pairwise distinct 
heads $a_1,\ldots, a_n$. If the family of all stable models of $P$
is $\{\{a_1\},\ldots,\{a_n\}\}$, then ${\overline P} =
CP[\{a_1,\ldots,a_n\}]$.
\end{lemma}
\proof Consider the program ${\overline P}$. Assume that it consists of
rules $r_1,\ldots,r_n$. Without loss of generality we will assume that
the head of $r_i$ is $a_i$, $1\leq i\leq n$. 

Observe that since $r_1$ is generating for $\{a_1\}$, the only
positive literal it may contain is $a_1$. So, assume that $a_1$ appears
positively in the body of $r_1$. Then, ${\overline P}^{\{a\}}$ contains
the rule $a_1\lar a_1$. Since all other rules in ${\overline P}^{\{a\}}$
have atoms different from $a_1$ in their heads, $a_1$ does not belong
to the least model of ${\overline P}^{\{a\}}$, a contradiction. Hence,
$r_1$ has no positive literals. By symmetry, all rules $r_i$ have no
positive literals in their bodies.

Next, observe that $r_1$ is generating for $\{a_1\}$ but not for any
other stable model $\{a_i\}$ ($i\not= 1$). Hence, all literals $\n(a_i)$,
$2\leq i\leq n$,
must appear in the body of $r_1$ and $\n(a_1)$ does not. Since $r_1$
has no redundant negative literals,
\[
r_1 = \ \ \ a_1\lar \n(a_2),\ldots,\n(a_n).
\]
By symmetry, it follows that ${\overline P} = CP[\{a_1,\ldots,a_n\}]$.
\halmos

To prove Theorem \ref{thm.main}, we establish the basis of induction in 
Lemma \ref{ind.b} and the induction step in Lemma \ref{ind.1}.

\begin{lemma}\label{ind.b}
Let $P$ be an extremal program with $n$, $2\leq n\leq 4$ clauses. 
Then, for some atoms $a$, $b$, $c$ and $d$:
\begin{enumerate}
\item if $n=2$, ${\overline P}= CP[\{a,b\}]$ $(= B(0))$,
\item if $n=3$, ${\overline P}= CP[\{a,b,c\}]$ $(= A(1))$,
\item if $n=4$, ${\overline P}= CP[\{a,b,c,d\}]$ $(= C'(1))$,
or ${\overline P}= CP[\{a,b\}] \cup CP[\{c,d\}]$ $(= C(1))$.
\end{enumerate}
\end{lemma}
\proof Let $P$ be an extremal program with $n$ clauses, $2\leq n\leq 4$. 
Since $P$ is extremal, $P$ has at least $n$ stable models (note that
$B(0)$ has 2 stable models, $A(1)$ has 3 stable models, and $C(1)$ and
$C'(1)$ have 4 stable models each).

Let $H$ be the set of heads of the rules in $P$. Then, each stable model
of $P$ is a subset of $H$, and all stable models of $P$ form an
antichain. If $|H|=1$, the largest antichain of subsets of $H$ has one
element. Thus, $|H|\geq 2$. 

Observe also that since $P$ is extremal, its rules contain no positive
redundant literals in their bodies (Lemma \ref{bounds.1}). Additionally, by 
the construction of ${\overline P}$, its rules contain no redundant negative 
literals, either. Hence, the rules of ${\overline P}$ are built of atoms
in $H$ only.

Assume first that $n=2$. Then, $|H|=2$, say $H=\{a,b\}$.
There is only one antichain of subsets of $H$ that has two elements:
$\{\{a\},\{b\}\}$. Hence, $P$ has two stable models: $\{a\}$ and
$\{b\}$. The assertion follows by Lemma \ref{aux}.

Assume next that $n=3$. If $|H|=2$, then the largest antichain
of subsets of $H$ has two elements, a contradiction (recall that $P$ has
at least three stable models). Hence, $|H|=3$, say $H=\{a,b,c\}$.
The program ${\overline P}$ has three rules, say $r$, $s$ and $t$, with
heads $a$, $b$ and $c$, respectively.

There are only two antichains of subsets of $H$ with three elements:\\
(1) $\{\{a,b\},\{a,c\},\{b,c\}\}$, and \\
(2) $\{\{a\},\{b\},\{c\}\}$.\\
Hence, the family of stable models of $P$ (and, hence, also of 
${\overline P}$) 
is either $\{\{a,b\},$ $\{a,c\},$ $\{b,c\}\}$ or $\{\{a\},\{b\},\{c\}\}$.

Consider the first possibility. Assume that rule $r$ contains a negative
literal. Clearly, rules $r$ and $s$ are generating
for $\{a,b\}$. Thus, the only negative literal that they may contain
is $\n(c)$. Reasoning in the same way, we find that the only negative 
literal that may be contained in the rules $r$ and $t$ is $\n(b)$, a
contradiction. Hence, $r$ and, by symmetry, $s$ and $t$ have no negative
literals. Thus, ${\overline P}$ is a Horn program and has exactly one
stable model, a contradiction.

It follows that the family of stable models of ${\overline P}$
is $\{\{a\},\{b\},\{c\}\}$. Now, the assertion follows by Lemma \ref{aux}. 

Finally, assume that $n=4$. If $|H|\leq 3$, the size of any antichain of
subsets of $H$ is at most 3. Since $P$ has at least 4 stable models,
$|H|=4$. Assume that $H=\{a,b,c,d\}$ and that ${\overline P}$
consists of rules $r$, $s$, $t$, and $u$ with heads $a$, $b$, $c$ and
$d$, respectively. 

Let $\cal A$ be an antichain 
consisting of 4 or more subsets of $H$. Clearly, $\cal A$ contains
neither $\emptyset$ nor $H$. Assume that ${\cal A}$ contains a
one-element subset of $H$, say $\{a\}$. Then, there are exactly two
possibilities for $\cal A$:\\
(1) ${\cal A} =\{\{a\},\{b\},\{c\},\{d\}\}$, and\\
(2) ${\cal A} =\{\{a\},\{b,c\},\{b,d\},\{c,d\}\}$.\\
In the first case, the assertion
follows from Lemma \ref{aux}. So, let us consider the second case.   
In this case, rule $r$ is not generating for any of the stable models
$\{b,c\}$, $\{b,d\}$ and $\{c,d\}$. Hence, $\{b,c\}$, $\{b,d\}$ and
$\{c,d\}$ are the stable models of ${\overline P}\setminus \{r\}$. This
is a contradiction. We proved above that no 3-rule program can have the
antichain $\{\{b,c\},\{b,d\},\{c,d\}\}$ as its family of stable models.

Next, assume that $\cal A$ contains a set with three elements, say
$\{a,b,c\}$. Then, there are exactly two possibilities for $\cal A$:\\
(1) ${\cal A} =\{\{a,b,c\},\{a,b,d\},\{a,c,d\},\{b,c,d\}\}$, and\\
(2) ${\cal A} = \{\{a,b,c\},\{a,d\},\{b,d\},\{c,d\}\}$. \\
Assume the first
case. Assume that at least one rule in ${\overline P}$, say $r$, has a
negative literal. Since $r$, $s$ and $t$ are generating for $\{a,b,c\}$, 
it follows that $r$ has exactly one negative literal, $\n(d)$. But then,
$r$ is not generating for $\{a,b,d\}$, a contradiction. Hence, $r$ and,
by symmetry, all the rules in ${\overline P}$ have no negative literals
in their bodies. Consequently, ${\overline P}$ is a Horn program and has
only one stable model, a contradiction.

Thus, assume that ${\cal A} = \{\{a,b,c\},\{a,d\},\{b,d\},\{c,d\}\}$.
Assume that $r$ has a negative literal. Reasoning as before, it follows
that $r$ has exactly one negative literal, $\n(d)$. But then, $r$ is not
generating for the stable model $\{a,d\}$, a contradiction. Hence, $r$
and, by symmetry, $s$ and $t$ have no negative literals in their bodies.
Assume that $u$ has a negative literal in its body, say $\n(x)$. Then, 
since $u$ is generating for $\{a,d\}$, $\{b,d\}$ and $\{c,d\}$, 
$x\notin \{a,d\}\cup \{b,d\}\cup \{c,d\}$, which is impossible. Hence, as
before, ${\overline P}$ is a Horn program and has only one stable model,
a contradiction.

The last case to consider is when $\cal A$ contains only sets consisting
of two elements. First, assume that some three sets in $\cal A$ contain
the same element, say $a$. Then  $\{a,b\}$, $\{a,c\}$ and $\{a,d\}$
are all in $\cal A$. Since $r$ is a generating rule for all three stable
models, it contains no negative literals and the only positive literal
it may contain in its body is $a$. Since facts do not belong to extremal
programs (Lemma \ref{bounds.1}), $a$ is in the body of $r$. Consequently, 
$a\lar a$ is in ${\overline P}^{\{a,b\}}$. Hence, $a$ is not in
the least model of ${\overline P}^{\{a,b\}}$, a contradiction.

The only remaining possibilities for $\cal A$ are\\
 (1) ${\cal A} = \{\{a,c\},\{a,d\},\{b,c\},\{b,d\}\}$,\\
 (2) ${\cal A} = \{\{a,b\},\{a,d\},\{b,c\},\{c,d\}\}$, \\
 (3) ${\cal A} = \{\{a,b\},\{a,c\},\{c,d\},\{b,d\}\}$. \\
They are isomorphic, so it is
enough to consider one of them only, say the first one.

Assume that $r$ has a positive literal in its body. Since $r$ is 
a generating rule for $\{a,c\}$ and $\{a,d\}$, it follows that  
$r$ has exactly one such literal, namely $a$. Hence, rule $a \lar a$
is in ${\overline P}^{\{a,c\}}$. Since no other rule in ${\overline
P}^{\{a,c\}}$ has $a$ as its head, $a$ is not in the least model
of ${\overline P}^{\{a,c\}}$, a contradiction. Hence, $r$ and, by
symmetry, all rules in ${\overline P}^{\{a,c\}}$ have no positive
literals in their bodies. 

Next observe that $r$ is generating for $\{a,c\}$ and $\{a,d\}$
and it is not generating for $\{b,c\}$ and $\{b,d\}$. Since it has no
positive literals in the body, it follows that $r = a\lar \n(b)$. 
By symmetry, clauses $b\lar \n(a)$, $c\lar \n(d)$ and $d\lar \n(c)$
are all in ${\overline P}$. Hence, ${\overline P} = CP[\{a,b\}]
\cup CP[\{c,d\}]$.
\halmos

Now, we will establish the induction step.

\begin{lemma}\label{ind.1}
Let $n$ be an integer, $n \geq 5$. Assume that every extremal program
with $2\leq n' < n$ rules and no redundant atoms is a $2,3,4$-program.
If $P$ is an extremal program with $n \geq 5$ rules and no redundant atoms
than:
\begin{enumerate}
\item $P$ contains no two rules with the same head
\item $P$ contains no atoms that appear only positively in the bodies of
the rules in $P$
\item $P$ contains no rules of the form $q\lar p$
\item $P$ is a $2,3,4$-program
\end{enumerate}
\end{lemma}
\proof Our assumption that every extremal program
with $2\leq n' < n$ rules and no redundant atoms is a $2,3,4$-program
implies that for every $n'$, $2\leq n' <n$, $s(n') = s_0(n')$.

\noindent
(1) Let $r = q \lar a_1,\ldots,a_k, \n(b_1),\ldots,\n(b_l)$ be a rule 
in $P$. Assume that there is another rule $r'$ with head $q$. From 
Lemma \ref{bounds.1} it follows that $k > 0$
or $l > 0$. Moreover, from Lemma \ref{bounds.1} we have that
that there is a rule $r''$ such that $q$ appears in the body of $r''$. 
Also, since there are no redundant rules in $P$, $r''$
is different than $r$ and $r'$.

If $q$ appears positively in the body of $r''$ then
$|P(q^-)| \leq n-3$. Since  $|P(q^+)| \leq n-2$, the inequality
(\ref{inq.f0b}) in Lemma \ref{bounds.1} and the 
inductive assumption imply that
\[
s(P) \leq s(P(q^+)) + s(P(q^-)) \leq 
s_0(n-2) + s_0(n-3) < s_0(n).
\]
So, P is not extremal. 

Assume then that $q$ appears negatively
in the body of $r''$. Now, $|P(q^-)| \leq n-2$, $|P(q^+)| \leq n-3$ and
we can show that $s(P) < s_0(n)$ in the same way as before.
Hence, $P$ contains no two rules with same head and (1) follows.

Therefore, for every atom $q$ which appears as a head in $P$, there is 
exactly
one rule with head $q$. We will denote this rule by $r(q)$.

\noindent
(2) Assume that $P$  contains an atom $q$ which appears only positively
in bodies of rules of $P$. There is a unique rule $r(q)$. Let
\[
r(q) = q \lar a_1,\ldots,a_l, \n(b_1), \ldots, \n(b_m)
\]
and $P'$ be the program obtained from $P$ by replacing every 
premise $q$ by the sequence 
$a_1,\ldots,a_l,$ $\n(b_1), \n(b_m)$. Then $|P| = |P'|$ and the programs 
$P$ and $P'$ have the same stable models. Also, $P'$ contains an atom which
never appears in a body of a rule in $P$. So, from Lemma \ref{bounds.1}
it follows that $P'$ is not extremal. Hence, $s(P) < s(n)$, a contradiction.

\noindent
(3) Assume that $P$ contains a rule of the form
$r = q \lar p$. Since there is only one rule 
in $P$ with head $q$, for every stable model $M$ of $P$, $q \notin M$ 
if and only if $p \notin M$.
Let $P'$ be the program obtained from $P$ by replacing every
premise $\n(q)$ by the premise $\n(p)$. Clearly,
$P$ and $P'$ have the same stable 
models. In addition, $P'$ contains an atom which does not
appear negated in $P'$. From part (2) of this proof, it follows that $P'$ is
not extremal. Consequently, since $P$ and $P'$ have the same number of 
rules and the same number
of stable models, $P$ is not extremal, contrary to the assumption.

\noindent
(4) Assume first that $P$ contains a rule $r$ of the form $q \lar \n(p)$.
Let $M\in \St(P)$. If $q\in M$, then $M\setminus\{q\} \in 
\St(P(r^+))$. If $q\not \in M$, then, $M\in \St(P(r^-))$
and $p\in M$. Hence, $M\setminus \{p\} \in \St((P(r^-))(r(p)^+))$
(recall that $r(p)$ is the unique rule in $P$ with $p$ as its head, cf. 
part (1) of the proof). Hence, 
\[
s(P) \leq s(P(r^+)) + s((P(r^-))(r(p)^+)).
\]

Observe now that $|P(r^+)|\leq n-2-\delta$, where $\delta$ is the number
of rules different from $r(p)$ and containing $\n(q)$ in the body.

Next, observe that $|(P(r^-))(r(p)^+)| \leq n-2-\epsilon$, where
$\epsilon$ is the number of literals in the body of $r(p)$ different than
$q$ and $\n(q)$.
Therefore,
\[
s(n) = s(P) \leq s(P(r^+)+s((P(r^-)(r(p)^+) 
\leq s(n-2-\delta) + s(n-2-\epsilon).
\]
If $\delta > 0$ or $\epsilon > 0$ then the
inequality \ref{inq.f0b} of Lemma \ref{inq.f0} and the equality 
$s(n')=s_0(n')$,
for $2\leq n'<n$, imply that 
$s(n) < s_0(n)$.
It follows that $\delta=0$, $\epsilon=0$
and both $P(r^+)$ and $P(r^-)(r(p)^+)$ are extremal. 
Moreover, since $\epsilon=0$, $r(p) = p \lar \n(q)$ ($P$ does not
contain redundant rules and rules of the form $p\lar q$).

Let $P' = P\setminus \{r,r(p)\}$.
Since $\delta = 0$, it also follows that there are no rules in $P'$
with $\n(q)$ in the body. By symmetry, it follows that
no rule of $P'$ contains $\n(p)$.

Assume now that there is a rule in $P'$, say $r'$, containing $q$
in its body. Again, let $M\in \St(P)$. If $q\in M$, then
$M\setminus \{q\}$ is a stable model of $(P(q^+))(p^-)$. Otherwise,
$M$ is a stable model of $P(p^+)(q^-)$. Since $|(P(q^+))(p^-)|\leq n-2$
and $|(P(p^+))(q^-)|\leq n-3$, 
\begin{eqnarray*}
s(P) &\leq& s(P(q^+)(p^-)) + s(P(p^+)(q^-)) \leq s(n-2) + s(n-3)\\
&=& s_0(n-2) + s_0(n-3) < s_0(n)\leq s(n),
\end{eqnarray*}
a contradiction. Hence, neither $q$ nor (by symmetry) $p$ appear
in $P'$. It is easy to see that $P'=P(r^+)$. Since $P(r^+)$ is extremal,
$P'$ is extremal. It follows by induction that $P'$ and, consequently,
$P$ are both $\{2,3,4\}$-programs. 

From now on, we will assume that every rule in $P$ has at least 2 literals
in the body. Assume that there is a rule $r$ in $P$ with a positive 
literal, say $a$, in its body. Since the body of $r(a)$ has 
at least two literals, $|P(a^+)|\leq n-3$. Since $r$ has $a$ in its body,
$|P(a^-)|\leq n-2$. It follows that $s(P)\leq s(n-3)+s(n-2) =
s_0(n-3)+s_0(n-2) < s_0(n) \leq s(n)$,
a contradiction. Hence, every rule in $P$ has only negative literals in its
body.  

Assume next that there is a rule $r$ in $P$ with $k\geq 4$ literals 
in the body. Let $q$ be the head of $r$. Then $|P(q^+)|\leq n-5$ 
and $|P(q^-)|\leq n-1$. Hence, $s(P) \leq s(n-5)+s(n-1) = 
s_0(n-5) + s_0(n-1) < s_0(n) \leq s(n)$, a contradiction.
It follows that every rule in $P$ has 2 or 3 literals in its body.

We will show now that $P$ is a $\{2,3,4\}$-program. To this end, we will 
consider two cases. First, we will assume that all rules in $P$ have exactly 3
negative literals in their bodies. Consider a rule $r$ from $P$, say $r$
is of the form:
\[
a\lar \n(b),\n(c),\n(d).
\]
Assume that the rules $r(b)$, $r(c)$, and $r(d)$ are of 
the following respective forms (by our assumption, each must have exactly
3 negative literals in the body):
\[
b\lar \n(x),\n(y),\n(z)
\]
\[
c\lar \n(x'),\n(y'),\n(z')
\]
\[
d\lar \n(x''),\n(y''),\n(z'').
\]
Assume that at least one of the atoms $x$, $y$, $z$, $x'$, $y'$, $z'$,
$x''$, $y''$ and $z''$ is not in $\{a,b,c,d\}$. Without the loss of generality,
we may assume that $x''\notin \{a,b,c,d\}$.

For a stable model $M$ of $P$, let $G_M$ denote the set of generating 
rules for $M$.
Then, we have the following four mutually exclusive
cases for $M$:
\begin{enumerate}
\item[(i)] $r(a)\in G_M$
\item[(ii)] $r(a)\notin G_M$ and $r(b)\in G_M$
\item[(iii)] $r(a)\notin G_M$, $r(b)\notin G_M$ and $r(c)\in G_M$, and
\item[(iv)] $r(a)\notin G_M$, $r(b)\notin G_M$, $r(c)\notin G_M$ and
$r(d)\in G_M$.
\end{enumerate}
If $r(a)\in G_M$ then by Corollary \ref{cor.r} $M\setminus\{a\}$ is a stable
model of $P(r(a)^+)$. Since $|P(r(a)^+)| \leq n-4$ 
the number of stable models for which (i) holds is bounded by $s(n-4)$.

Similarly, by considering $P(r(b)^+)$ and $P(r(c)^+)$ we have that
the number of stable models  for which (ii) or (iii) hold
is bounded, in each case, by $s(n-4)$. 

Consider  $P(r(d)^+)$. Since $x''\notin 
\{a,b,c,d\}$, the number of stable models for which (iv) holds is 
bounded by 
$s(n-5)$.  Hence, $s(P) \leq 3s(n-4) +s(n-5)$.
Lemma \ref{f.incr} implies that  $s(P) < 4s(n-4)$. Using the inductive
assumption and, twice, the inequality \ref{inq.f0a} of Lemma \ref{inq.f0} 
we have that
$4s(n-4)=4s_0(n-4) \leq s_0(n)$. So, $s(P) < s_0(n)\leq s(n)$.
This is a contradiction. Consequently, all atoms appearing in the negated 
form in the bodies of the rules $r(b)$, $r(c)$ and $r(d)$ 
belong to $\{a,b,c,d\}$.  Hence, $\{r(a),r(b),r(c),r(d)\} = CP[\{a,b,c,d\}]$. 

Let us now observe that none of $\n(a)$, $\n(b)$, $\n(c)$ and $\n(b)$
appears in 
\[
P\setminus \{r(a),r(b),r(c),r(d)\}.
\]
Indeed, if, say $\n(a)$,
appears in the body of a rule $r(q)$, where $q\notin\{a,b,c,d\}$, then
one can show that $s(P) \leq s(n-5)+s(n-1) = s_0(n-5) +s_0(n-1) <
s_0(n) \leq s(n)$, a contradiction.

Since $s(P) \leq s(P(a^+)) + s(P(a^-)) \leq s(n-4) +s(n-1) = s_0(n-4)
+ s_0(n-1) \leq s_0(n) \leq s(n)$, it follows that $P(a^+)$ is extremal 
and that $P(a^+) = P\setminus
\{r(a),r(b),r(c),r(d)\}$. Consequently, $P\setminus\{r(a),r(b),r(c),r(d)\}$
is a $\{2,3,4\}$-program. Thus, $P$ is a $\{2,3,4\}$-program.

To complete the proof we need to consider one more case when
$P$ contains a rule, say $r(a)$, 
with exactly 2 negative literals in the body. Let us assume that
\[
r(a) = a \lar \n(b),\n(c)
\]
Let us also assume that $r(b)$ has literals $\n(x)$ and $\n(y)$ in its
body (and, possibly, one more) and that $r(c)$ has literals
$\n(x')$ and $\n(y')$ (and, possibly, one more) in its body.
If $r(b)$ or $r(c)$ has three negative literals in its body or if 
at least one of $x$, $y$, $x'$ and $y'$ is not in $\{a,b,c\}$, reasoning
as in the previous case we can show that $s(P) \leq 2s(n-3) + s(n-4) 
= 2s_0(n-3) +s_0(n-4) < 3s_0(n-3)$.
Corollary \ref{cor-4} implies that $3s_0(n-3) \leq s_0(n)\leq s(n)$.
Hence, $s(P) < s(n)$.
This is a contradiction.
Hence, $\{r(a),r(b),r(c)\} = 
CP[\{a,b,c\}]$. Moreover, again reasoning similarly as before, we can show
that none of $\n(a)$, $\n(b)$ and $\n(c)$ occurs in $P\setminus
\{r(a),r(b),r(c)\}$. Hence, $s(P) \leq s(P(a^+)) + s(P(a^-)) \leq
s(P(a^+)) + 2 s_0(n-3)\leq 3s_0(n-3)\leq s_0(n) 
\leq s(n)$. It follows that $P(a^+)$ is extremal.
Moreover, $P(a^+) = P\setminus \{r(a),r(b),r(c)\}$. Consequently,
$P\setminus \{r(a),r(b),r(c)\}$ is a $\{2,3,4\}$-program and, thus, so is
$P$.
\halmos

We can now complete the proof of Theorem \ref{thm.main}. Let $P$ be an
extremal program. Then, by Lemmas \ref{ind.b} and \ref{ind.1}, ${\overline P}$
is a 2,3,4-program. Thus, by Corollary \ref{cor-4}, $P \in {\cal E}_n$. 
Consequently, $s(n) = s_0(n)$. \halmos

\bibliographystyle{alpha}

\end{document}